\documentclass[aps,amsmath,amssymb,reprint,superscriptaddress,twocolumn,showpacs,longbibliography]{revtex4-1}

 \usepackage{graphicx}
 \usepackage{amsmath}
 \usepackage{amsfonts}
 \usepackage{amssymb}
 \usepackage{pstricks}
 \usepackage{psfrag}
 \usepackage{epsfig}
 \usepackage{changes}
 \usepackage[ansinew]{inputenc}
\usepackage{mathptmx}
\usepackage{helvet}
\usepackage{textcomp}
\usepackage{ulem}

\begin{document}


\title{Optimal  cell transport in straight channels and  networks}



\author{Alexander Farutin}
\affiliation{Univ. Grenoble Alpes, LIPHY, F-38000 Grenoble, France}
\affiliation{CNRS, LIPHY, F-38000 Grenoble, France}
\author{Zaiyi Shen}
\affiliation{Univ. Grenoble Alpes, LIPHY, F-38000 Grenoble, France}
\affiliation{CNRS, LIPHY, F-38000 Grenoble, France}
\author{Gael Prado}
\affiliation{Univ. Grenoble Alpes, LIPHY, F-38000 Grenoble, France}
\affiliation{CNRS, LIPHY, F-38000 Grenoble, France}
\author{Vassanti Audemar}
\affiliation{Univ. Grenoble Alpes, LIPHY, F-38000 Grenoble, France}
\affiliation{CNRS, LIPHY, F-38000 Grenoble, France}
\author{Hamid Ez-Zahraouy}
\affiliation{Laboratoire de Mati\`ere Condens\'ee et Sciences Interdisciplinaires, Faculty of Sciences, Mohammed V University of Rabat, Morocco}
\author{Abdelilah Benyoussef}
\affiliation{Laboratoire de Mati\`ere Condens\'ee et Sciences Interdisciplinaires, Faculty of Sciences, Mohammed V University of Rabat, Morocco}
\author{Benoit Polack}
\affiliation{Laboratoire d'H\'ematologie, CHU, Grenoble, France}
\affiliation{TIMC-IMAG/TheREx, CNRS UMR5525, Universit\'e Grenoble Alpes, Grenoble, France}
\author{Jens Harting}
\affiliation{Helmholtz Institute Erlangen-N\"urnberg for Renewable Energy (IEK-11), Forschungszentrum J\"ulich, F\"urther Strasse 248, 90429 N\"urnberg, Germany}
\affiliation{Department of Applied Physics, Eindhoven University of Technology, P.O. Box 513, 5600MB Eindhoven, The Netherlands}
\affiliation{Faculty of Science and Technology, Mesa+ Institute, University of Twente, 7500 AE Enschede, The Netherland}
\author{Petia M. Vlahovska}
\affiliation{Engineering Sciences and Applied Math, Northwestern University, Evanston 60208, USA}
\author{Thomas Podgorski}
\affiliation{Univ. Grenoble Alpes, LIPHY, F-38000 Grenoble, France}
\affiliation{CNRS, LIPHY, F-38000 Grenoble, France}
\author{Gwennou Coupier}
\email[]{gwennou.coupier@univ-grenoble-alpes.fr}
\affiliation{Univ. Grenoble Alpes, LIPHY, F-38000 Grenoble, France}
\affiliation{CNRS, LIPHY, F-38000 Grenoble, France}
\author{Chaouqi Misbah}
\email[]{chaouqi.misbah@univ-grenoble-alpes.fr}
\affiliation{Univ. Grenoble Alpes, LIPHY, F-38000 Grenoble, France}
\affiliation{CNRS, LIPHY, F-38000 Grenoble, France}


\date{\today}

\begin{abstract}
 Flux of rigid or soft particles (such as drops, vesicles, red blood cells, etc.) in a channel is a complex function of  particle concentration, which depends on the detail of induced dissipation and suspension structure due to hydrodynamic interactions with walls or between neighboring particles. Through 2D and 3D simulations and a simple model  that reveals the contribution of the main characteristics of the flowing suspension, we discuss the existence of an optimal volume fraction for cell transport and its dependence on the cell mechanical properties. The example of blood is explored in detail, by adopting the commonly used  modeling of red blood cells dynamics.  We highlight the complexity of optimisation at the level of a network, due to the antagonist  evolution of local volume fraction and optimal volume fraction with the channels diameter. In the case of blood network, the most recent results on the size evolution of vessels along the circulatory network  of healthy organs suggest that the red blood cell volume fraction (hematocrit) of healthy subjects is close to optimality, as far as transport only is concerned. However, the hematocrit  value  of patients suffering from diverse red blood cells pathologies
 may strongly deviate from optimality.
\end{abstract}


\maketitle


\section{Introduction}

Lab-on-a-chip technologies allow for high throughput analysis or manipulation at the level of single particles, in particular biological cells or droplets. They are now part of scientific and industrial landscape. While most concepts have been initially developed on dilute suspensions, the possibility of manipulating more concentrated suspensions is now explored \cite{Wyatt2015,Ley2016,Karthick2017}. The expected gains in throughput have to be evaluated in light of the potential decrease of the process efficiency, due to cell-cell interactions \cite{VERNEKAR2015}, screening of the controlling external field (e.g. in acoustophoresis \cite{Karthick2017}), etc... In addition, increasing substantially particle concentration will eventually lead to an increase of the viscosity which will often lead to a decrease in the particle flux, far before non-reversible events like clogging occur \cite{Jensen13}.

This points to the question of the existence or not of an optimal volume fraction of particles that would optimize their transport for given boundary conditions like pressure difference at the inlet and outlet  of a channel network, as considered here.

This question naturally arises also in blood vascular  networks which have a  complex architecture, and  where oxygen is delivered to tissues via the red blood cell (RBC) hemoglobin. About $40-45\%$ of  blood volume is composed of RBCs (in the macrocirculation). Given its medical importance, e.g., in blood transfusion,  the existence  of an optimal hematocrit (optimal volume percentage of RBCs)  has been extensively discussed in the physiology and hemorheology literature \cite{Hedrick1986,Linderkamp1992,Wells1991,Stark2012,Birchard1997,Barbee1971,Lipowsky1980a,shepherd82}.

Optimality in living systems is a subtle question to be tackled with care. In addition to oxygen transport, blood has many other biological functions  such as thermoregulation, lymph production, immune response, or carbon dioxide, nutrients and waste transport. Yet, it is still worth asking whether RBC transport (and as a corollary oxygen transport) is optimum or not and, more importantly,  what is the impact of  cell mechanical properties  or network architecture characteristics on optimal  transport. Understanding these questions  may help understanding situations of hypoxia in relation with pathological evolution of these parameters, or provide hints towards  developing artificial blood. \\

We consider in this paper  blood as a reference example.  We shall first explore the impact of variations in the properties of red blood cells on the overall transport property (cell flow rate). More precisely, we investigate   the role of cell mechanical properties, that of  reduced volume  --more or  less deflated cell-- with incompressible membrane, as well as the role of membrane  	 (the case of capsules). In a second part, we will analyze  the effect of  network geometry on the cell flow rate. { Finite-size effects of the cells} plus the existence of cross-stream migration (lift)  of hydrodynamic origin generally lead to non-uniform distribution of the flowing particles \cite{philipps92,hudson03,katanov15,rivera16,Qi17}, which in turn leads to a decrease of volume fractions in smaller channels (called  F\aa hr\ae us effect for blood \cite{Fahraeus1929,Popel2005}), and uneven distribution of cells in a network of small channels, due to peculiar phase separation effects at the level of bifurcations \cite{barber08,doyeux11,shen16,roman16}. This points to the fact  that a global optimality at the level of a network is probably not the consequence of an optimization at each level, but most likely as a non local  compromise involving different levels of vessel branches.

In the case of blood flow, we find that optimality cannot be reached in all the vessels of the circulatory system at the same time.  For straight channels it is found that the optimal hematocrit (RBC volume fraction) matches the actual hematocrit in  large arterioles, in the range 100-200 $\mu$m (for humans the arteriole range  is $10-200 \mu m$ \cite{Klabunde2012}). For a vessel network (mimicking the real vasculature), we find that the  cell flux is dominated by these arterioles and larger vessels, and that the {hematocrit} (as defined in routine blood tests) of a healthy subject is close to optimality. \\

Solving blood distribution problems in silico is challenging and highly non-trivial because of the complex nature of blood flow: it involves  many deformable RBCs moving and interacting  in blood vessels with complicated  geometry \cite{Abkarian2008,Petia_review,Guido2009b,Abreu2014,Li2013}. Recent advances in computational power and novel simulation approaches, however, are beginning to make the task tractable and  the bottom-up  simulations of blood flow (by explicitly accounting for  blood elements) are becoming realistic \cite{Li2013}.
 Computational approaches based on continuum formulations \cite{Kraus1996,Cantat1999b,Pozrikidis2005,Bagchi2007,Dodson2008,Lac2007,Veerapaneni2009,Le2010,Zhao2010,Boedec2011,Salac2011,Lai2,Farutin2014b,rivera16} or particle-based models \cite{Noguchi2005a,Dupin2007,Clausen2010,Fedosov2011,Kaoui2011,Krueger2013, Pivkin2008} have  been   successfully applied to  simulations of the flow of single and many RBCs. These simulations have revealed that many  phenomena in blood flow such as platelet and leukocyte margination, plasma skimming, clustering \cite{Chien1987,Pries1992,Popel2005,pries96,McWhirter2009,Tomaiuolo2012,rivera16} have a purely mechanistic origin. For example, plasma skimming  and  hematocrit decrease in the microcirculation (F{\aa}hr{\ae}us  effect) \cite{Popel2005} arise from a cross-streamline migration of RBCs due to local flow perturbation by the deformable RBC  \cite{Olla1997,Cantat1999b,seifert1999}. Many experiments have pointed to and continue to reveal plethora of effects on the individual  and collective behaviors of RBCs under flow \cite{Fischer894,dupire12,Guido2009b,grandchamp13,Lanotte:2016,shen16}.

We investigate stationary blood flow properties  using both numerical simulations and analytical studies covering a wide volume fraction range. While in the arterioles the pulsatile nature of flow has not been fully damped yet, the associated Womersley number is of order 0.05 \cite{fung97}, therefore at each moment the velocity profile is similar to that of the  stationary case. In the simulations we use two models for cells: (i) a 3D model accounting for both bending energy and in-plane shear elasticity, and solved by the Lattice Boltzmann method (LBM). This 3D model will be also used to consider capsules with extensible membranes - while RBC ones are not ;  (ii) a 2D model based on the LBM  which accounts for membrane bending elasticity that shows the same trends, offering a faster tool to explore wider ranges of parameters. 
These simulations are completed by a discussion based on a simple model that incorporates the most important features of the rheology and structure of a suspension, which can be used to anticipate the impact of the variations of cell or network parameters.

\begin{figure*}[t]
\includegraphics[width=\textwidth]{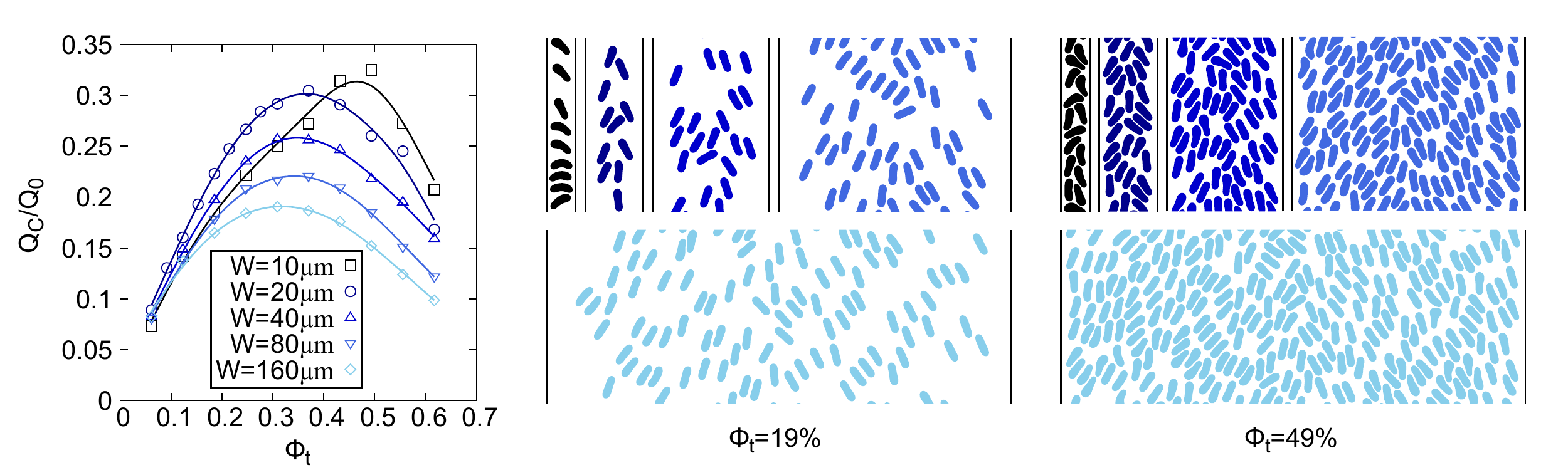}
\caption{\label{fig3} (2D simulations) Left pannel: Normalized cell flow rate as a function of volume fraction  for channels with different widths. Full lines are guide for the eyes. 
$C_a=0.9$ and reduced area is $\nu_{2D}=0.7$. Centre and right pannel: snapshots of the suspension for two different volume fraction in the five considered channels.}
\end{figure*}

 \begin{figure}[t]
\includegraphics[width=0.4\textwidth]{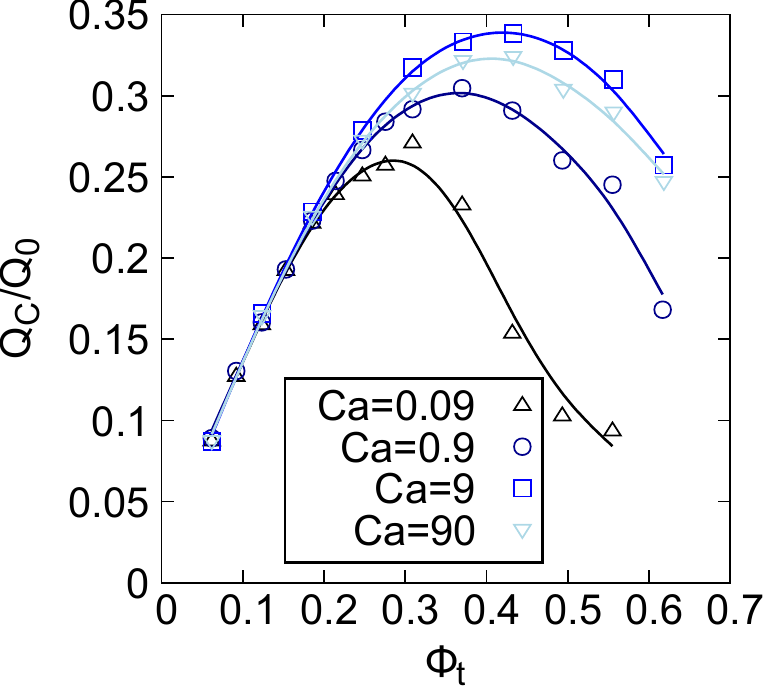}
\caption{\label{fig2}   (2D simulations)  Normalized cell flow rate as a function of volume fraction  for a channel width $W=20 \mu$m and different  $C_a$. Reduced area is $\nu_{2D}=0.7.$}
\end{figure}

\section{Materials and Methods}

We represent cells as either a contour endowed with bending energy immersed in 2D fluid, or as a surface having both
bending energy and shear elasticity in a 3D fluid. The simulation uses a 2D \cite{Kaoui2011} and 3D \cite{Krueger2011} Lattice Boltzmann method.  The bending energy is given by $E=(\kappa/2) \int H^2 dA$ with $H$ the mean curvature, $dA$ the arclength in 2D or area in 3D, and $\kappa$ the bending rigidity modulus. In 3D, the membrane has  additionally a  shear elastic energy written as
$\mu_s(I_1^2 + 2I_1-I_2)/12+  \kappa_\alpha I_2^2/12$,
where $\mu_s$ is the shear elastic modulus and $\kappa_\alpha$ is the area dilation modulus.
$I_1$ and $I_2$ are the in-plane strain invariants (see \cite{Krueger2011}).

We impose a pressure difference $\Delta p$ between the inlet and outlet of a straight channel of length $L$ and width $W$ for several values of volume fraction $\Phi_t$. The  channel widths range from $W=2R$, where $R$ is the cell effective radius,  up to $W=64 R$, thus covering the range of in-vitro applications and the capillary to arteriole scale for the blood network. In the absence of cells the flow is of a Poiseuille type. Periodic boundary conditions are considered in the longitudinal direction and also in the third direction in 3D. In 2D {(3D)} the volume fraction in the channel, called tube volume fraction $\Phi_t$, is defined as $\Phi_t= N A /{\cal A}$ ($N V/{\cal V})$  where $N$ is the number of cells in the channel, $A$  the area of cell and ${ V}$ the volume, whereas ${\cal A}$ and ${\cal V}$ designate the total area and volume of the channel. $\eta$ is taken as $1.2$ m.Pa.s (plasma viscosity), and the enclosed fluid within cells is taken to have the same viscosity. The precise values of the viscosity contrast have little influence on the results, as shown below. We define the capillary numbers (which are a measure of the flow strength over the cells mechanical resistance) associated with bending and shear elastic modes as
$C_a= \eta \langle\dot{\gamma}\rangle R^3/\kappa$ and $C_{s}= \eta \langle\dot{\gamma}\rangle R/\mu_s$, with $\langle\dot{\gamma}\rangle=2U_{max}/W$,  where $U_{max}$ is the maximum velocity in the channel in the absence of cells. As a reference, we have taken   $\kappa \simeq 3 \; 10^{-19}\;$ J and $\mu_s\simeq 4 $ \textmu N/m, close to that known for RBCs \cite{Suresh2006}. We define in 2D $R =\sqrt{A/\pi}$, and in 3D $R=[3V/(4\pi)]^{1/3}$ as the typical cell radius. For a healthy RBC we have $R\simeq 2.7 \mu m$.

We will explore values from 0.09 to 90  for  $C_a$  in 2D and from  0.2 to 0.9  for $C_s$ in 3D. This latter range corresponds to that found for red blood cells, considering typical vessel diameters of human capillaries and arterioles (about 5-10 $\mu$m and 10-200 $\mu$m respectively \cite{Klabunde2012})   and typical maximal velocities in capillaries and arterioles ($1-10 $mm/s) \cite{KOUTSIARIS2010202}.

We define (in 2D) the reduced area $\nu_{2D}\equiv (A/\pi)/(p/2\pi)^2$ (with $p$ the perimeter and  $A$ the enclosed area) and the reduced volume (in 3D) $\nu\equiv [V/(4\pi/3)]/[A/4\pi]^{3/2}$.  For a healthy RBC, $\nu$ is reported to be in the range $\simeq 0.6-0.64$ \cite{linderkamp83,Fung93}.  Here, we chose as a reference $\nu=0.64$, as in other simulation papers of the literature \cite{Li2013,cordasco14,Lanotte:2016}. Higher values  will also be explored. For blood,  several diseases (spherocytosis, ellipsocytosis)  correspond to high value of $\nu$, close to one. $ \nu=1$ plus incompressible membrane also corresponds to the case of a suspension of hard spheres.
In 2D, we choose $\nu_{2D}=0.7$, which is the reduced area of the section along the main axis of a prolate spheroid of reduced volume $0.6$. 

Finally, we varied $\kappa_\alpha/\mu_s$ between 1.25 and 125 to explore the effect of area dilation. The value 125 is our reference case and is large enough to preserve membrane area locally and corresponds to red blood cells, while lower values allow to consider the case of what is usually called elastic capsules.


\begin{figure*}
\centering
\includegraphics[width=0.9\textwidth]{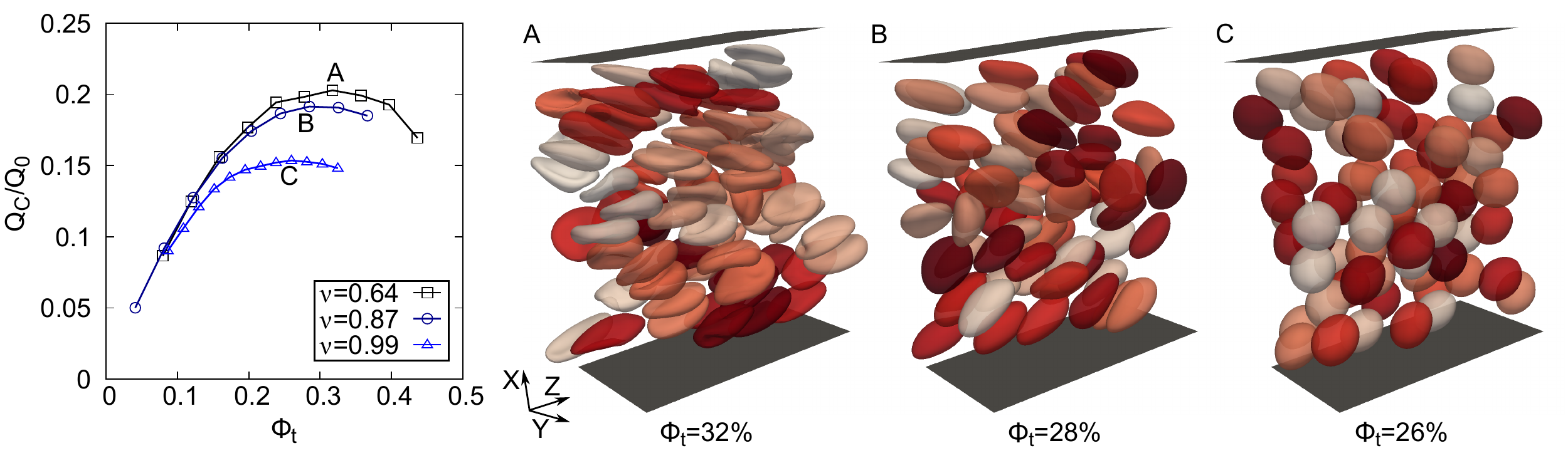}
\caption{\label{zaiyi3D}(3D simulations)  Left: The normalized cell flow rate as a function of tube volume fraction $\Phi_t$ for different reduced volumes.
Viscosity contrast is 1 and the channel width $W$ is $40 \mu$m$=14.8 \;R $,  the channel length along the flow direction is about $11.7\; R$, and is equal to $6.2\; R $ in the orthogonal direction.  $C_s=0.18$. A, B and C are snapshots of suspension configurations for the three  different reduced volumes shown for the corresponding optimal volume fraction.}
\end{figure*}

\begin{figure}
\centering
\includegraphics[width=0.5\textwidth]{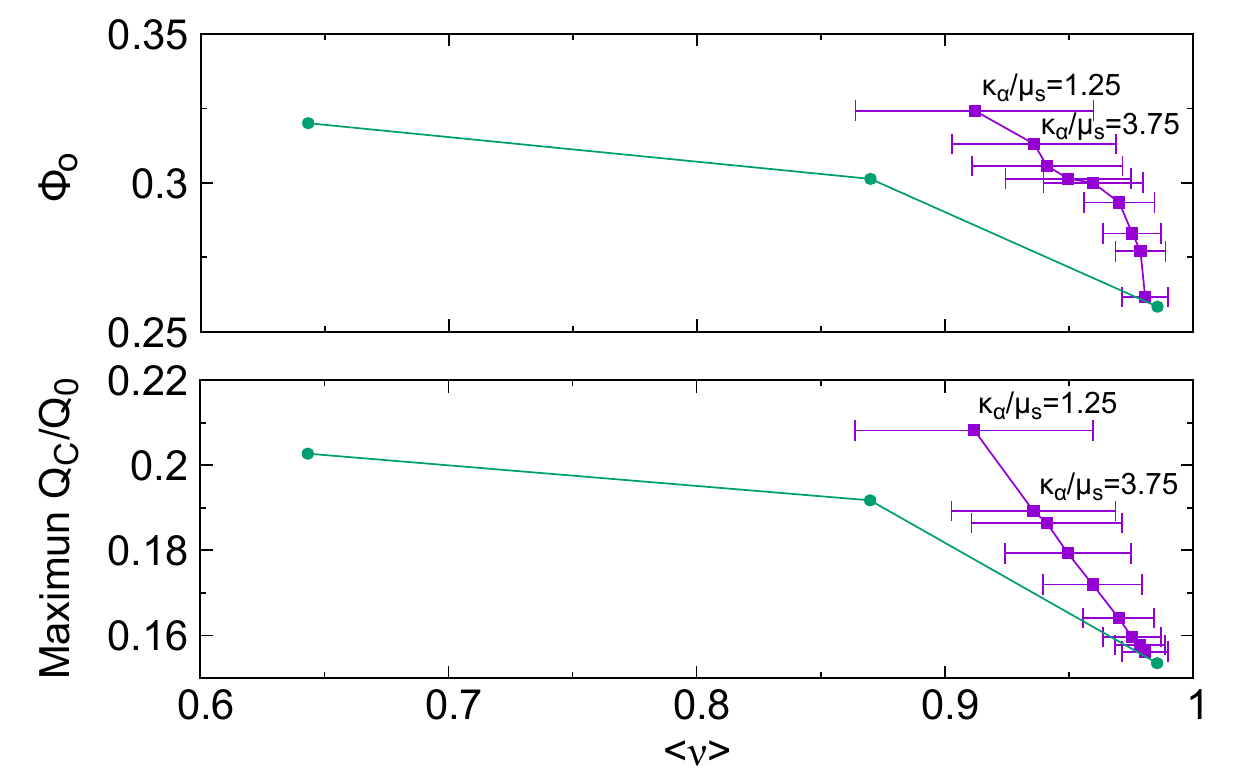}
\caption{\label{fig6} (3D simulations)   Green line: Optimal flux and optimal volume fraction as a function of the actual reduced volume (3D simulations) for incompressible membranes (from Fig. \ref{zaiyi3D}). Violet lines: the same simulation with stretchable  capsules. For each simulation the compressional elasticity $\kappa_\alpha/\mu_s$ has been varied in order to allow for different average reduced volume $\langle  \nu \rangle$. Each capsule experiences different shear stresses within the channel, and has thus different  surface area (and thus reduced volume); the horizontal bars provide the distribution width of reduced volumes within the same suspension. $C_s=0.18$, as in Fig. \ref{zaiyi3D}.}\end{figure}

\section{Optimal flux in a straight channel}



Throughout the paper, we focus on  the  cell flow rate $Q_{C}$ normalized by the flow rate of the cell-free fluid $Q_0$ under same pressure gradient.  In the simulations, the  flow rate $Q_{C}$ is calculated by counting the number of cells that cross a given section in the channel, averaged over long runs.


\subsection{Simulation results}

\paragraph{Role of flow characteristics.}

We investigate first the effect of flow strength (measured by the capillary number) and channel width. These simulations are performed in 2D.

Figure \ref{fig3}  shows that $Q_{C}/Q_0$ exhibits a maximum as a function of volume fraction $\Phi_t$. It illustrates the effect of  channel width on the corresponding optimal volume fraction $\Phi_o$.  The results show that $\Phi_o$ varies between 0.3 in large enough channels ($W=160\mu m$) to about 0.4 in smaller channels   ($W=10\mu m$). The corresponding cell flux also increases when the channel gets narrower. This shift of the cell flow rate curves is related to the increase of the effective viscosity with the channel diameter. This effect is known for blood as the F{\aa}hr{\ae}us-Lindquist effect \cite{Popel2005}.  It stems from the fact that the thickness of the cell-free layer near the walls depends only weakly on the channel diameter, as it is essentially the result of local interactions.  Therefore, the relative contribution of this low-viscosity layer to the effective viscosity of the whole fluid becomes less and less important as the channel diameter increases. 

Figure \ref{fig2}  shows that the maximum of  $Q_{C}/Q_0$  is quite sensitive to the flow strength $C_a$: the optimal volume fraction  $\Phi_o$ increases from 0.3 to more than 0.4 over  3 orders of magnitude of increase in $C_a$.  The corresponding cell flux increases in the same proportion. These results can be considered as a consequence of the increased cell deformation that makes the suspension less viscous (shear thinning behavior).

\paragraph{Role of cell properties.}

We now turn to the effect of modifications of the cell properties, which is explored through 3D simulations. Figure \ref{zaiyi3D} shows that a spherical shape of the cells lowers both the optimal $\Phi_o$ and the maximum cell flow rate. For example, when the reduced volume $\nu$ passes from 0.64  to 0.99  the maximal carrying capacity drops by about 25 $\%$. This is attributed to the fact that
cells are less deformable because of nearly spherical shape reducing their ability
to squeeze and accommodate high $\Phi_t$. Moreover, cross-stream migration is suppressed and the cell-free layer diminishes, leading to increased flow resistance.

When membranes are incompressible, reduced volume determines the ability of cells to deform. This ability can also be lost by an alteration of the mechanical property (e.g. shear elasticity) of the membrane. For example, in sickle cell and malaria diseases, the RBC elastic modulus can be significantly higher (up to about three times higher \cite{Suresh2006}) than within healthy subjects. An increase of the elastic modulus is equivalent to a decrease of the capillary number. Our data in Fig. \ref{fig2} show that a reduction of the capillary number leads to a significant decrease of the cell flow rate.

The role of membrane compressional elasticity on the maximum cell flow rate is explored in Figure \ref{fig6}. The green data correspond to incompressible membranes. To explore the effect of variations of the area dilation modulus $\kappa_\alpha$ (violet data),  we have selected  an initial reduced volume { $\nu=1$} (the same for all cells) and performed several simulations, each time with different elastic properties ($\kappa_\alpha/\mu_s$ between 1.25 and 125). Each cell within the same uniform suspension  (i.e.  all cells have the same properties)   experiences different shear stress and thus will be more or less stretched. The average actual reduced volume within the suspension is shown as a  filled square, whereas the horizontal bars show the distribution of reduced volume for each simulation (i.e.  for a given stretching elastic coefficient).
  One sees that a moderate extensibility $\kappa_\alpha/\mu_s   \approx 1.25$, allowing an average decrease of the reduced volume of the capsules from $\langle \nu \rangle=1 $ to $\approx 0.9$, leads to a maximum flow rate $\max(Q_C/Q_0)\approx 0.2$ equivalent to that of healthy incompressible red blood cells, even though the capsules are quasi-spherical.
  Interestingly, the cell flow rate is more important for stretchable capsules than for non stretchable cells having the same reduced volume.  The ability to deform leads to better accommodation to the flow stress, with different shapes within the suspension.

 \subsection{Analytical model}

We develop an analytical model that captures the computational observations for the dependence of $Q_{C}$ on tube volume fraction $\Phi_t$. This analytical model provides means for a quick estimate of the optimal volume fraction  and insight into the mechanisms that control the phenomenon, as explored by our simulations.
We adopt a continuous two-fluid model for  suspension flow. Such an approach has already proved to give interesting insights for other problems related to blood flow \cite{secomb87,cokelet91,sharan01}. Here, we consider an inner core containing a homogeneous suspension of volume fraction $\Phi$ surrounded by a cell-free layer of thickness $e$, flowing in a tube of radius  $R_0=W/2$. For given tube volume fraction $\Phi_t$, $ \Phi= \Phi_t \times (\frac{R_0}{R_0-e})^2$.

The cell flux is
\begin{equation}
Q_{C}(\Phi_t)= \Phi \int_0^{R_0-e} v (\Phi,r) 2 \pi r  dr,\end{equation}
and the total flux is   given by
\begin{equation}
Q_{T}(\Phi)= \int_0^{R_0}  v (\Phi,r) 2 \pi r  dr.
\end{equation}
 For a given pressure gradient (flow strength), the Stokes velocity profiles $v(\Phi,r)$ in the outer cell-free annulus and in the core can be evaluated. 
  Assuming that the core fluid is a homogeneous, dense suspension of concentration $\Phi$, its relative viscosity $\bar\eta\equiv \eta(\Phi)/\eta_0$  can be estimated  from the  Krieger-Dougherty  law  $\bar\eta=(1-\Phi/\Phi_m)^{-[\eta] \Phi_m}$, where $\Phi_m$  is  the maximum fraction,  $[\eta]$ is the intrinsic viscosity, and $\eta_0$ is the viscosity of the cell-free fluid \cite{Krieger1959}. This law was chosen since it is simple enough to allow discussion on how it should evolve with the cell properties, but accurate enough to describe viscosity evolution with cell concentration: Quemada law, a particular case of Krieger-Dougherty  law, describes well the bulk behavior of blood \cite{sharan01}. To first order, $e$ is independent of $w$, at least for $w>15$ \textmu m \cite{fedosov10}, and is mainly determined by the balance between the lift flux and the diffusive flux. This can be seen on the central and right pannels of  Fig. \ref{fig3}. Due to the linear relationship between the diffusion constant and local volume fraction \cite{dacunha96,grandchamp13}, we assume that the thickness of the outer cell-free layer decreases linearly as volume fraction increases, an assumption  which is validated in the literature \cite{fedosov10,shen16,rivera16}. Consequently, we set $e=e_0(1-\Phi_t/\Phi_m)$ where $e_0$ is a constant characterizing the cell-free layer at low volume fraction. Strictly speaking, this model is not valid for very low fraction where $e$ is expected to converge to $w/2$. Despite the oversimplification, this model captures the main features as long as we focus on large enough $\Phi$ so that hydrodynamic interactions among cells  counterbalance the effect of walls. This is indeed the range where the optimal volume fraction is expected. Note that very few interaction events are sufficient to balance the lift force, because the latter strongly decreases with distance to the wall \cite{grandchamp13,olla97,abkarian02,coupier08}.
 Finally,
\begin{equation}
\begin{aligned}
\frac{Q_{C}}{Q_0} &= \frac{Q_{C}}{Q_T(\Phi=0)} \\
&=\frac{\Phi \big(R_0-e\big)^2\Big(R_0^2+\big(2 \bar\eta (\Phi)-1\big)(2 eR_0-e^2)\Big)}{\bar\eta (\Phi) R_0^4},
\end{aligned}
\end{equation}

where $e$ and $\Phi$ are the functions of $\Phi_t$ given above.

\begin{figure}
\centering
\includegraphics[width=0.5\textwidth]{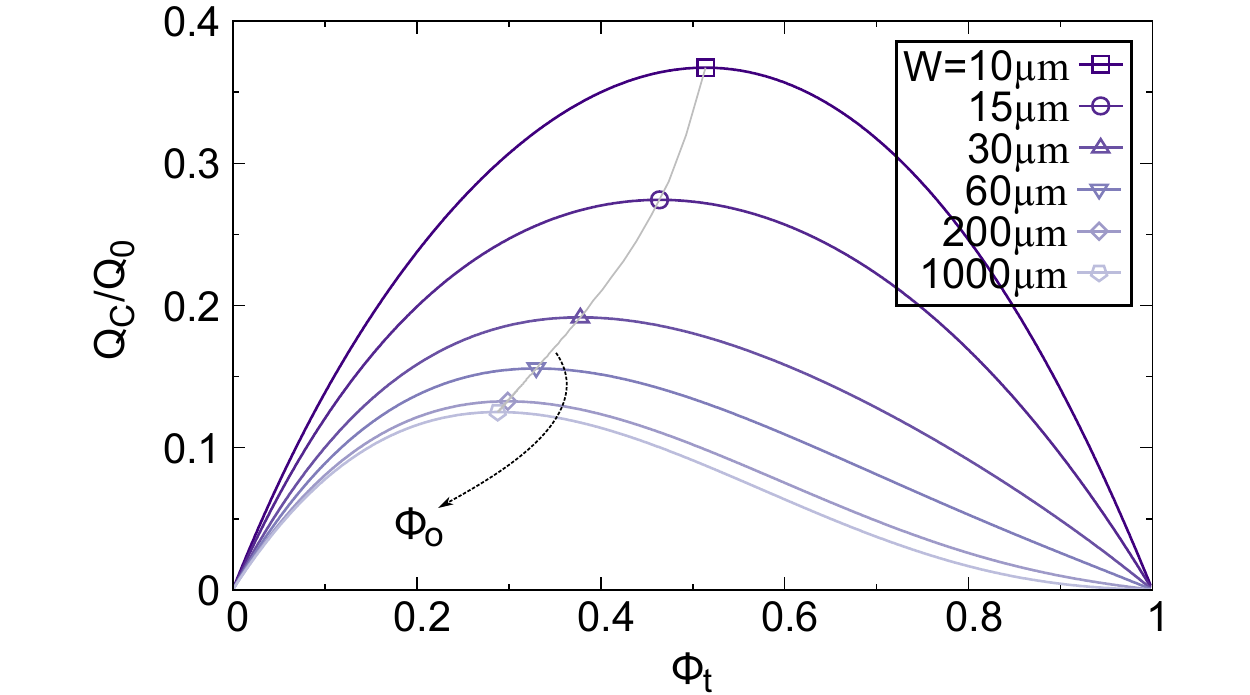}
\caption{\label{fig4} The particle flow rate as a function of tube volume fraction for  channels with different diameter  calculated from our minimal model. We took   $e_0=2 \mu$m, $[\eta]=5/2$ as for rigid  spheres and for RBCs in physiological conditions \cite{vitkova08},  and $\Phi_m$ is chosen as equal to 1 considering that the cells are deformable and can be squeezed.
}
\end{figure}

Figure  \ref{fig4}  illustrates the  prediction of this reduced model.
  As in simulations, it is seen that the optimal volume fraction increases with the confinement, and so does the corresponding cell flux.  Had we assumed a fixed cell-free layer, independent of $\Phi_t$, we would then have missed the fact (data not shown, already shown in Ref. \cite{Stark2012}) that optimal volume fraction increases with confinement.

\begin{figure}[h]
\centering
\includegraphics[width=8cm]{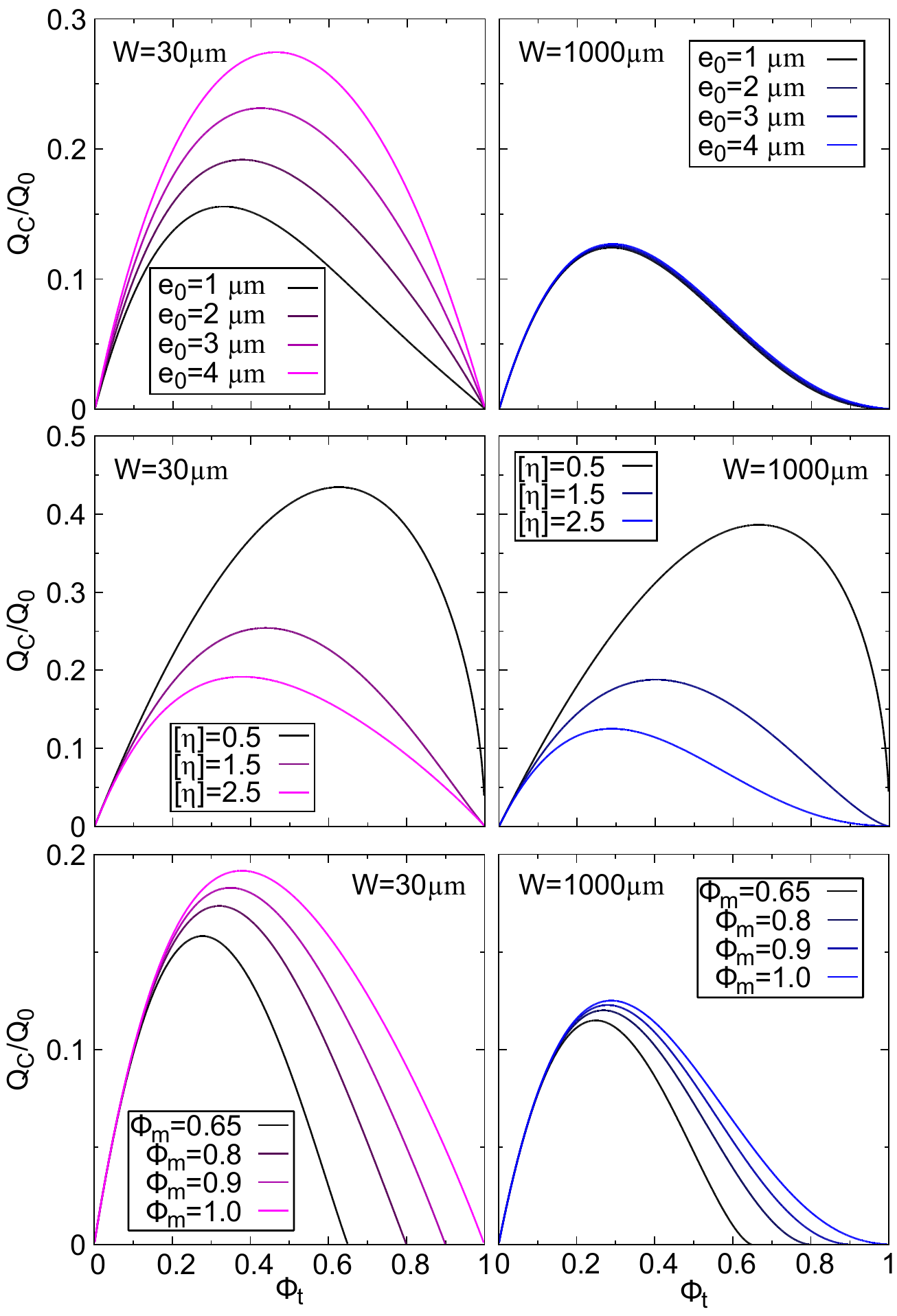}
\caption{\label{model} RBC flux according to our minimal model. The reference configuration is $W=30$ \textmu m (left) or  $W=1000$ \textmu m (right) , $e_0=2$ \textmu m, $[\eta]=5/2$ and $\Phi_m=1$. Top panel: $e_0$ is varied ; middle panel: $[\eta]$ is varied ; bottom panel: $\Phi_m$ is varied.}
\end{figure}

We  now exploit further the results of this simplified model to help proposing scenarios for the evolution of the optimal volume fraction with some cell or flow properties: see Fig. \ref{model}.
The evolution of the optimal volume fraction with the reduced volume, seen on Fig. \ref{fig6} (obtained numerically), can be understood by using our minimal model: if one considers less deflated cells, the intrinsic viscosity $[\eta]$ is not expected to vary too much since it is equal to 5/2 for spheres and close to that value for RBCs \cite{vitkova08}. On the other hand, inflating cells  leads to a decrease of the maximal packing fraction $\Phi_m$, and to a decrease of the size of the cell free layer  $e_0$ \cite{olla97,callens08,coupier08}. Both effects  lead to  to a decrease of the optimal volume fraction and of the associated cell flux  (Fig. \ref{model}, bottom and top panel), in agreement with Fig. \ref{fig6}.

 Upon an increase of the capillary number, one expects the size of the cell free layer to increase because of the increase of cell deformation and the resulting lift force \cite{grandchamp13}, and the intrinsic viscosity to decrease, characterizing generic shear-thinning behavior of capsule suspensions. As seen  in Fig. \ref{model}, top and middle panel, both effects lead to an increase of the optimal volume fraction, in agreement with  Fig. \ref{fig2}.

\begin{figure}
\centering
\includegraphics[width=0.4\textwidth]{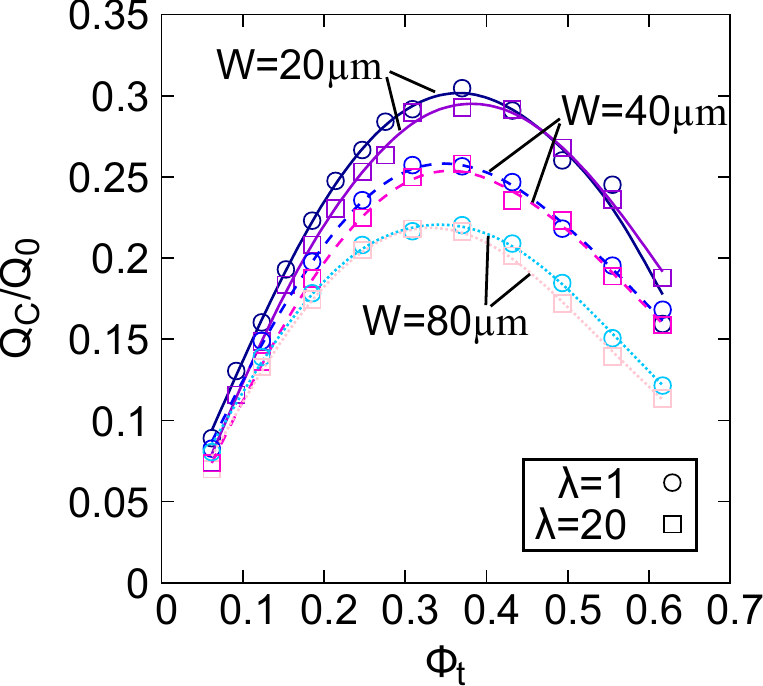}
\caption{\label{lambda} (2D simulations)  Normalized cell flow rate as a function of volume fraction  for channels with different widths and two different viscosity ratio $\lambda$ between the inner fluid and the external fluid.
$C_a=0.9$ and reduced area is $\nu_{2D}=0.7$.}
\end{figure}

Another situation  which is classically explored in the literature regarding  RBC deformation is to act on  the viscosity of the suspending fluid \cite{vitkova08,grandchamp13, shen16,dupire12,fischer13}. By increasing the external viscosity the lift forces are expected to increase \cite{grandchamp13,bureau17} hence the cell free layer thickness also increases \cite{shen16}. The impact on the rheology is less trivial. Contrary to drops for which the intrinsic viscosity is a monotonous function of external viscosity, the subtle interplay between inner fluid dynamics and the deformations allowed by the presence of the membrane lead to a non monotonous function in the case of cells with membranes \cite{vitkova08,Ghigliotti2010}. Both for structuring \cite{shen16} and for rheology \cite{vitkova12} effects, the effect of decreased external viscosity is strongly reduced by a concentration increase. In particular, cells that would tumble in a dilute suspension tends to tank-tread when the cell concentration increases, as they would do in a more viscous suspending fluid. All in all, we can see in Fig. \ref{lambda}, that our 2D simulations show that in reality the flux does not depend much on the viscosity ratio between the inner and the outer fluid.

\section{Optimal flux in a network}.

The preceding results can be used to infer the flux of particles in a network. However, one must keep in mind that volume fraction is in general not a conserved quantity in a network (say from one branch of the network to another),  or even in a straight tube of varying section.  As a preliminary, we discuss this point first before presenting our results of flow rate in a network.

\subsection{Tube and reservoir volume fraction}
\label{sec:fahr}
Starting from a large reservoir filled with a suspension of volume fraction $\Phi_r$, one generally observes that the fraction decreases once the suspension enters a smaller channel.  This effect, which is called F\aa hr\ae us effect \cite{Fahraeus1929} for blood, results from the tendency of cells to accumulate in the center of the vessel (due to the wall-induced migration), acquiring thus a higher mean velocity than that of the suspending fluid and leading to a diluted flow in small channels: $\Phi_t<\Phi_r$.  By contrast,  in large channels, the relative effect of the wall-induced migration is  negligible and the cells are homogeneously distributed within the channel section, implying $\Phi_t\simeq \Phi_r$.

By considering the conservation of particle flux, the reservoir volume fraction can be calculated from the results of a simulation or of the minimal model in a given tube: if $\langle V_{C}\rangle$ denotes the mean velocity of cells and $\langle V_T\rangle$ that of the whole suspension (cells plus the suspending fluid) in the tube, the total flow rate of cells across a section $A$ of the tube is given by $Q_{C}=\langle V_{C}\rangle A \Phi_t$ and that of the whole  suspension is given by $ Q_T=\langle V_T\rangle A$. The ratio of  $Q_{C}$ to $Q_T$ provides the volume fraction of cells found at the entrance or at the exit,  $\Phi_r$. We have thus the following relation
\begin{equation}
\label{reservoir}
\Phi_r=\frac{Q_{C}}{Q_T}= \Phi_t \frac{\langle V_{C}\rangle}{\langle V_T\rangle}
\end{equation}

\begin{figure}
\centering
\includegraphics[width=0.5\textwidth]{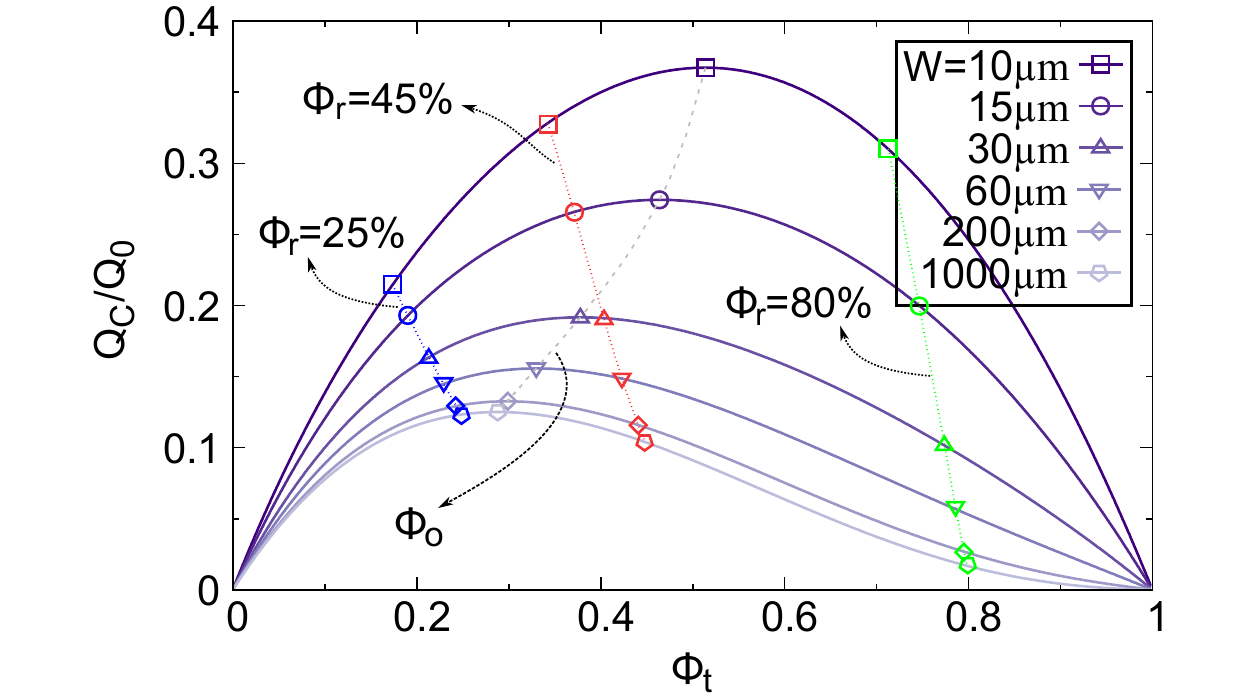}
\caption{\label{fig4-bis} The particle flow rate as a function of tube volume fraction from our minimal model (same data as in Fig. \ref{fig4}). In addition, the red line indicates the $\Phi_t$ value corresponding to a reservoir volume fraction $\Phi_r$ of 45\%  for  the corresponding model. Blue and green lines correspond to high $\Phi_r$ (80\%), and low $\Phi_r$ (25 \%).
}
\end{figure}

\begin{figure}
\centering
\includegraphics[width=0.5\textwidth]{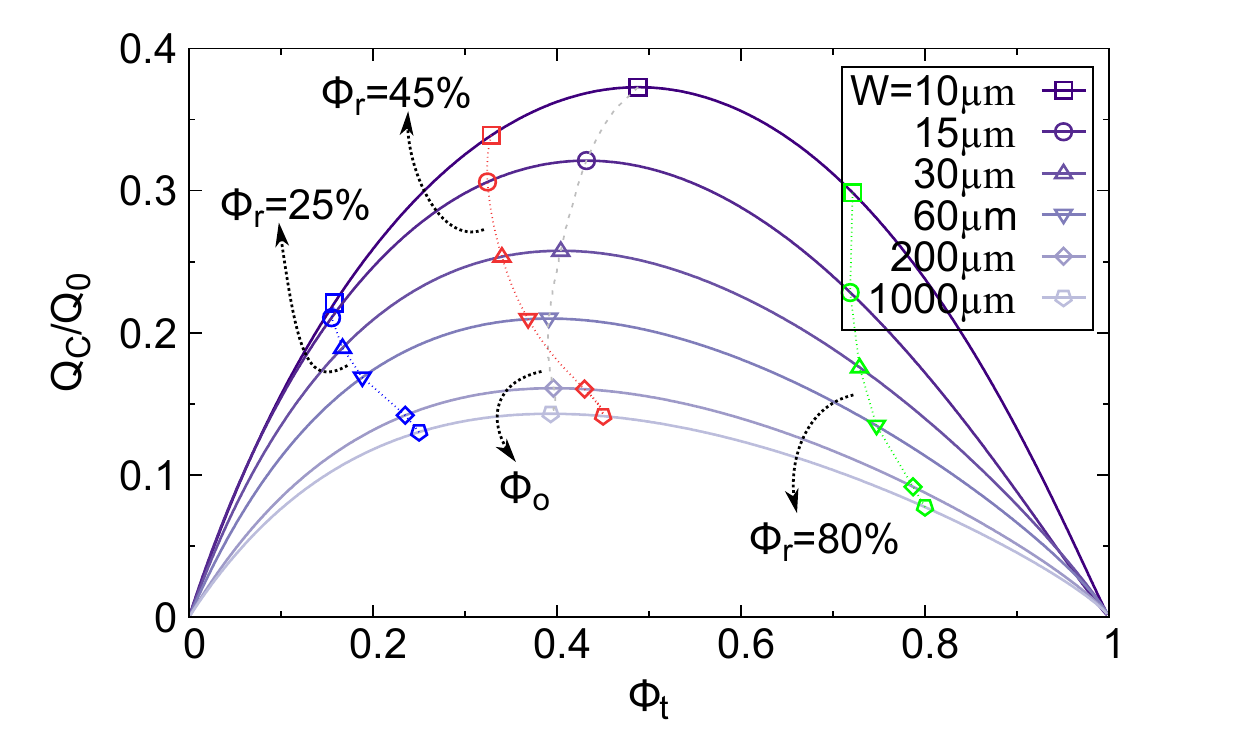}
\caption{\label{figp}The RBC flow rate as a function of tube hematocrit for  channels of different diameters calculated
from Pries empirical model \cite{Pries1992};
The red line indicates the $\Phi_t$ value corresponding to a reservoir hematocrit of 45\%.  Blue and green lines correspond to high $\Phi_r$ (80\%, polycythemia), and low $\Phi_r$ (25 \%, anemia).}
\end{figure}

In Fig. \ref{fig4-bis}  the $\Phi_t$ values for $\Phi_r=0.45$ are indicated for our minimal model, and one clearly sees first  that the discrepancy between  $\Phi_t$ and $\Phi_r$ increases with the confinement, and second that in a network fed by a given value of $\Phi_r$, the flux can be suboptimal in some branches and superoptimal in others. 

This is true also in the vascular network: Pries \textit{et al.}  considered in Ref. \cite{Pries1992} a fit of experimental data to propose empirical relationship between the effective viscosity of blood in a cylindrical tube of diameter $W$ and the tube hematocrit (=volume fraction in red blood cells) $\Phi_t$, as well as between this tube hematocrit and the reservoir hematocrit $\Phi_r$.  Both functions $\eta_{\mbox{p}}(\Phi_t)$ and $\Phi_r(\Phi_t)$  are given in  \cite{Pries1992}. The relative RBC flow rate at constant pressure drop is then given by $Q_{C}/Q_0=\Phi_r(\Phi_t)\times \frac{\eta_{\mbox{p}}(\Phi_t=0)}{\eta_{\mbox{p}}(\Phi_t)}$.

Figure \ref{figp} shows this relative RBC flow rate as a function of tube hematocrit for different channel diameters. The existence of an optimal hematocrit is highlighted, which increases with the confinement. The same holds for the  maximum RBC flow rate. Note that the value of this maximum as well as its position are well captured by our minimal model (Fig. \ref{fig4}).

The usually referenced hematocrits, measured in medical blood tests, are in the ranges  0.34-0.45 and 0.39-0.49  for women and men respectively \cite{troussard14}, and correspond to values encountered in macrocirculation. The red points in Fig. \ref{figp} show the tube hematocrit that would result from $\Phi_r=0.45$ reservoir (systemic) hematocrit.

Patients suffering from low or high level of hematocrit (anemia and polycythemia, respectively) are clearly far from optimality. Indeed,  if the reservoir hematocrit is small enough ($\Phi_r=0.25$), as happens in severe anemia disease -- blue line in Fig. \ref{figp} --, or large enough ($\Phi_r=0.8$) as is the case in polycythemia vera -- green line in Fig. \ref{figp} --, then the tube hematocrit does not cross any optimal value when varying the tube diameter. In the anemia disease case the hematocrit remains suboptimal, while it remains superoptimal in the polycythemia disease. In the first case the attained maximal flow rate in arterioles is  about 25\% lower than in healthy subjects, while the situation is more severe in the second case where the reduction can attain about 60\%.

Back to the healthy case, one notices that the tube hematocrit $\Phi_t$ corresponding to $\Phi_r=45\%$ is optimal in large channels until  $W=60$ $\mu$m, which corresponds to the medium diameter range of human arterioles. This is interesting inasmuch as two thirds of the oxygen is  known to be delivered in the arteriolar trees \cite{Gellis} before reaching capillaries.  On the contrary, in capillaries, the tube hematocrit is  suboptimal  for a reservoir hematocrit of 45\%. This shows that optimality in a network can probably not be the result of local optimality everywhere in the network.  The relative contribution of each branch to the overall particle flux must therefore be estimated.

\begin{figure}
\centering
\includegraphics[width=0.48\textwidth]{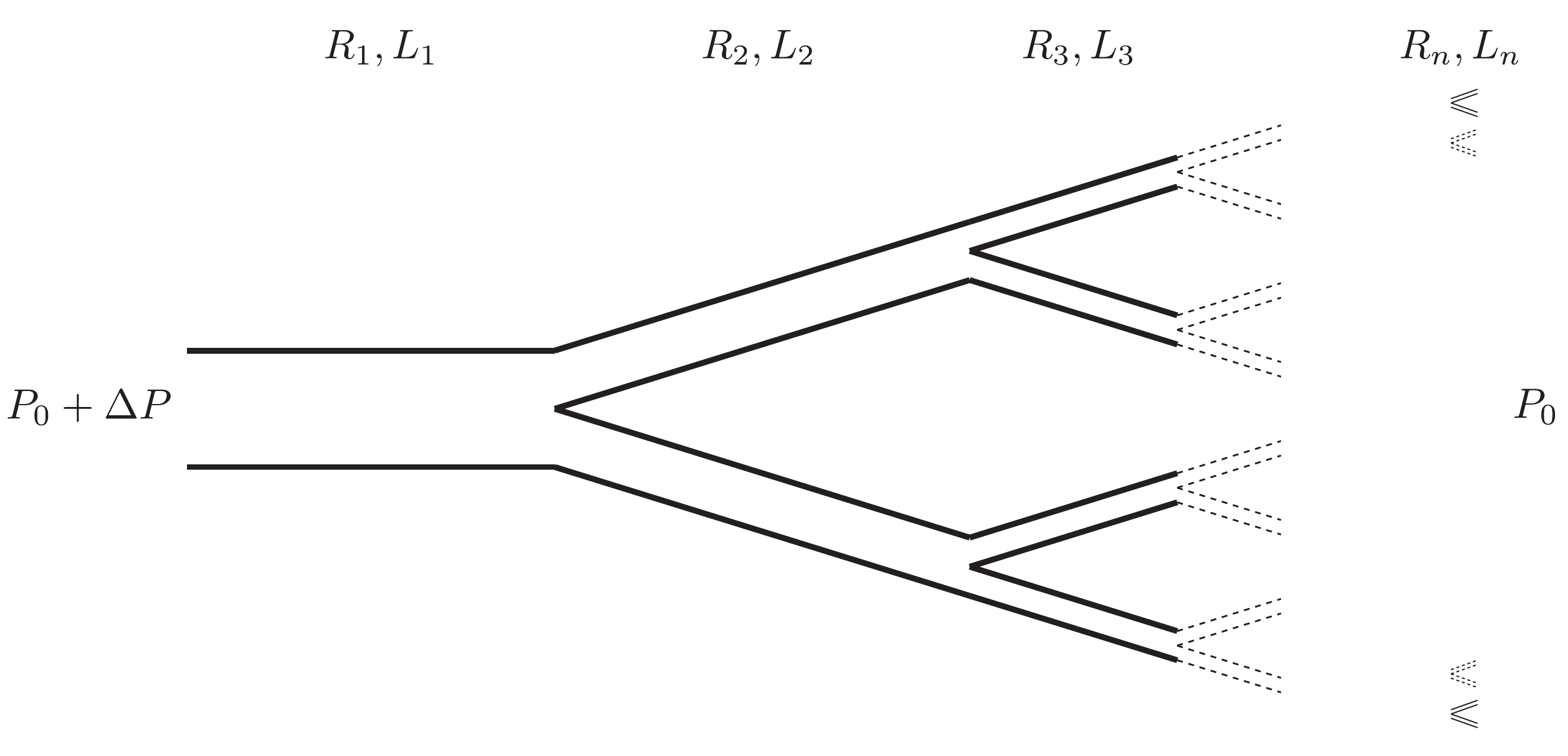}
\caption{\label{figdessin}Sketch of the model network considered in section \ref{sec:net}.}
\end{figure}

\subsection{Particle flux in a network}
\label{sec:net}
We consider a simplified network consisting of $N$ levels enumerated as $i=1...n$ (see Fig. \ref{figdessin}). Level 1 consists of one straight cylindrical channel. Each channel of level $i$ has radius $R_i$ and length $L_i.$ It divides symmetrically into two channels of level $i+1$. At each step, the cell and plasma fluxes also split symmetrically, so that $\Phi_r$   is a conserved quantity
(we recall that $\Phi_r$ is defined in Eq. \ref{reservoir} and is not to be confused with the actual hematocrit in a given branch of the network, which is the tube hematocrit). We consider a fixed pressure drop $\Delta P$ between the inlet of the level-1 channel and the outlets of the level-$n$ channels.  Assuming  bifurcation junctions do not contribute to pressure drop, one straightforwardly obtains:
  \begin{equation} Q_{C} =\Phi_r Q_T=\Phi_r \Delta P \times\big\{\sum_{i=1}^n 2^{-i+1}\frac{8 L_i \eta(\Phi_r,R_i)}{\pi R_i^4}\big\}^{-1}\label{eqnetwork}.\end{equation}

It is seen that the contribution of each channel depends on the detailed evolution of lengths and radii along the network. In addition,  as already discussed, $\eta(\Phi_r,R_i)$ is generally an increasing function of  $R_i$, therefore $\eta(\Phi_r,R_i)/R_i^4$ may depend on $R_i$ in a complex way. We now illustrate this complexity for the case of blood flow.

Recently, the relationships between radii $R_i$ and lengths $L_i$ for living organisms have been
  discussed in detail using a statistical analysis based on 3D imaging in human subjects \cite{newberry15}. Adapted to our case of dichotomous and symmetric branching, these relationships can be written as $R_i^{1/a}=2 R(i+1)^{1/a}$ and $L_i^{1/b}=2 L(i+1)^{1/b}$, where $a$ and $b$ are exponents that can be extracted from fitting of in-vivo data.
Using the recursive formulae $R_i=2^{-a(i-1)}R_1$ and $L_i=2^{-b(i-1)}L_1$ we find:
  \begin{equation}Q_{C} =\Phi_r \frac{\pi R_1^4 \Delta P}{8 L_1} \times\big\{\sum_{i=1}^n 2^{(4a-b-1)(i-1)}\eta(\Phi_r,R_i)\big\}^{-1}\label{eqnetworkscaled}.\end{equation}

The factor $2^{(4a-b-1)(i-1)}$ thus expresses the relative contribution from level $i$. According to \cite{newberry15}, a value for $a$ in the range $1/3-1/2$ yields a good description of real data, with $a=1/3$ for small vessels ($R_i<1$ mm) and $a=1/2$ for large vessels. An exponent-based scaling for the lengths is not as strongly supported by the data, since  different measurements yield  exponent $b$ in the range $0.17-1.40$. However, it is interesting to observe that for $a$ and $b$ within the aforementioned ranges, the exponent $4a-b-1$ can be either negative or positive, indicating that the major contributions can come from large or from small vessels depending on the exponent sign. Since the effective viscosity significantly depends on the tube radius only when the radius is smaller than about 1 mm \cite{Pries1992}, we focus on the case $a=1/3$, which is also consistent with  the  Murray's law \cite{Murray1926}. In that case, a realistic network geometry is likely to fall within a category where  $4a-b-1=1/3-b<0$ (since this is consistent with most available data for $b$), that is, a configuration where the flux would be regulated by large vessels.

\begin{figure}
\centering
\includegraphics[width=0.5\textwidth]{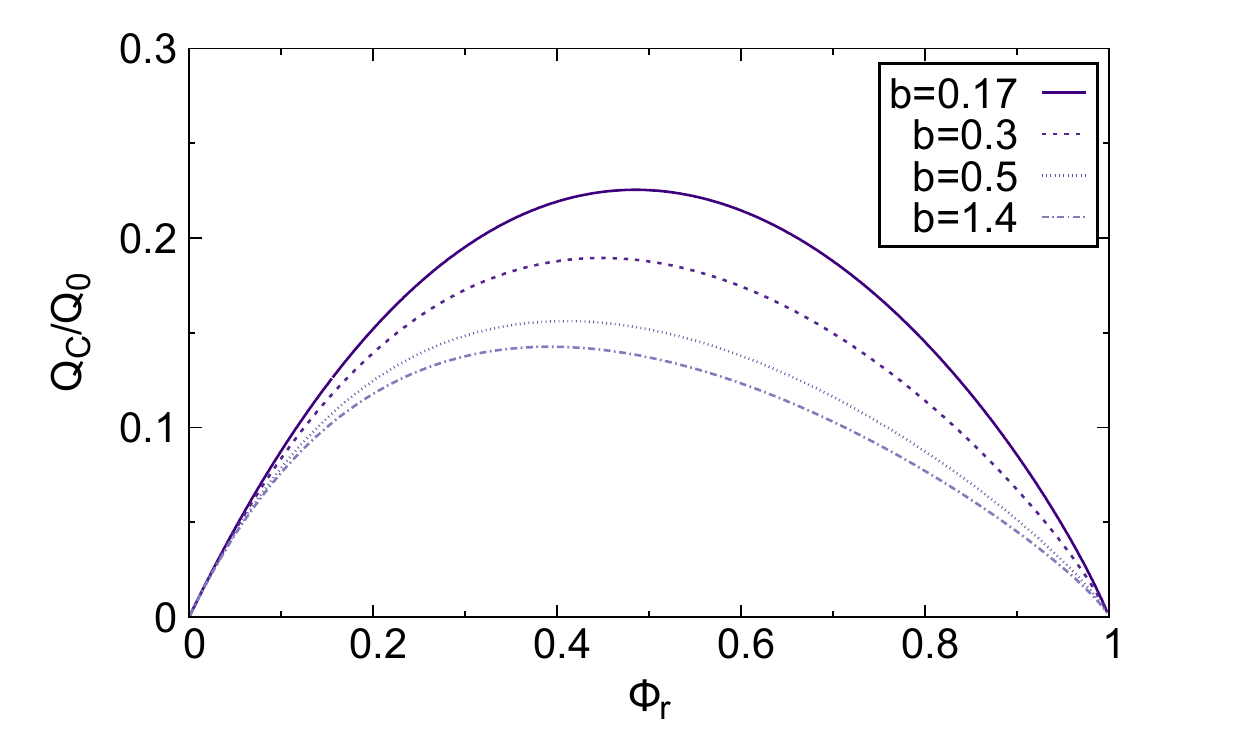}
\caption{\label{fignetwork} The RBC flow rate as a function of reservoir hematocrit for a model network (see text) for different values of $b$ extracted from literature and empirical rheology law given  in Ref. \cite{Pries1992}. The curves are similar to those for large channels that would be obtained from Fig. \ref{figp} by considering $\Phi_r$ instead of $\Phi_t$. Small values of $b$ give more weight to the smallest vessels, hence an increase of the optimal hematocrit. Still, the contribution of large vessels is predominant.}
\end{figure}

In order to be more quantitative and include the dependency of the viscosity with the vessel radius, we consider in  Fig. \ref{fignetwork} a network starting from a channel of about $1$ mm with successive bifurcations down to 5 $\mu m$, and plot the RBC flow rate as a function of reservoir hematocrit, as given by Eq. \ref{eqnetworkscaled}.
 Except for small $b$ values, the curves {are quantitiveley closer to those obtained  for large vessels than for small vessels (see Fig. \ref{figp})}, confirming that the dominant contribution arises from largest vessels of the network. This would point to the fact that optimization of oxygen transport capacity may be regulated  upstream of the microcirculation, close to the transition zone between small arteries and arterioles. Caution is however necessary since the situation is less clear for $a=1/2$, for instance.

This result also implies that  the  optimal hematocrit $\Phi_0$ is around  known physiological values.  However, given the dispersion of experimental data, the possibility for a positive exponent $4a-b-1=1/3-b$ is not to be excluded, in which case narrow channels  would have more weight than larger vessels. A systematic analysis of the vessel size distribution in organs is needed before drawing more conclusive answers. In addition, the regularly branched network considered here is not  the rule. 
 For example, and  as pointed out in Ref. \cite{karshafian03}, it has been shown that the network of vessels of a tumor, which grows much more rapidly than that of normal tissues, is chaotic and entangled, in contrast to the hierarchical branching pattern  adopted here.
More complex modelling for the geometry can be developed to account for this richness \cite{stamatelos14,rieger15,peyrounette18}. Handling this complexity and the different scales requires to pay specific attention to the computational feasibility and accuracy, which may lead to the development of more complex flow models through specific assumptions which can be validated by our approach. For instance, in \cite{peyrounette18}, the inhomogeneities and resulting nonlinearities due to complex cell partitioning in the capillary bed are explicitly neglected.

Indeed, while experimental studies of tube flow \cite{Albrecht1979,Fahraeus1929,Barbee1971}
yielded minimal values for the ratio $\Phi_t/\Phi_r$ of about 0.7 for tubes of 20 $\mu m$ and 10 $\mu m$ diameters, in agreement with Pries law,  in vivo studies \cite{Sarelius1982,Lipowsky1980a,Klitzman1979} have shown that this ratio can fall in the range 0.3-0.5  in the microcirculation.  These measurements have suggested that this is dependent on network topography and local flow rates. Asymmetric partitioning at the level of bifurcations can lead to important hematocrit variations between capillaries \cite{pries96}.  According to our network modeling, these variations should likely  have no strong impact on the issue of optimisation, as they occur in channels that  are expected to play a minor 
role on the flux regulation. However, asymmetric partitioning of cells or asymmetry in the network geometry may have a-priori an impact on the overall resistance of the capillary bed, and thus have a potential influence on optimality.

However, 
we expect this impact to be relatively small on the basis of the following argument. Let us consider two parallel branches originating from a bifurcation. We can calculate explicitly the effective resistance of this sub-network by using  Pries \textit{et al.} viscosity law \cite{Pries1992} and an adequate cell splitting model \cite{pries90}, which gives the cell flux ratio between the two branches as a function of their diameters ratio and the total flow rates ratio, in the case of upstream flow in stationary condition. If the mean diameter and the mean length between the two branches are kept constant and equal to the values given by the network scalings considered above, one finds that the resistance is always smaller than 110 \% of that of the symmetric case; note that in most cases, an even smaller value is found.  

Cell splitting is controlled in a subtle way by the geometry and the rheology law, so that giving a simple intuitive  argument about the overall behavior is not an easy task. However, some key features emerge from the following argument. First, we note that the conductance of the considered system is the sum of the conductances of the two branches. Conductance is proportional to the radius to the power 4, to the inverse of the branch length and to the inverse of the viscosity, which is a quasi-linear function of hematocrit in the range of interest. It is therefore a convex function of each of the 3 parameters  (length, diameter, hematocrit) taken separately, which implies that any asymmetry in the cell splitting  or in the geometry will mainly result in an increase of the conductance (and decrease of resistance and thus minor impact on optimality). Care must be taken, however. Since hematocrit repartition depends on the geometry, the different parameters are not independent and  the convexity argument may not be entirely sufficient. Still it serves as a reasonable guideline.

Another point is worth being mentioned. In general, in a given branch of the microvasculature, the RBCs distribution may not be centered, by lack of long enough distance between two successive bifurcations;
for instance all cells may be close to one wall. In that case the partitioning will be quite different from the case of fully centered distribution, discussed above. For a given geometry, the convexity argument mentioned above clearly shows that the resistance will be lower than in the homogeneous distribution. Typically, it will be controlled by that of the branch that receives fewer cells.

This discussion suggests that even in non-idealized cases, the role of capillaries in the determination of the overall resistance and flux in a network is secondary compared to large vessels. Of course, detailed numerical simulations are necessary to fully conclude about the impact of asymmetries, beyond this simplified case of 2-branch network.

\section{Conclusion}

2D and 3D numerical simulations  provided information on the behavior of cell flow rate as a function of cell volume fraction in a straight channel. Based on a  minimal model we highlight  the cell free layer as a key element to understand the variation of the transport capacity of  channels with diameter. This leads to the conclusion that the transport capacity of a whole network depends on its precise architecture, since two antagonist  effects enter  into play when traversing channels from large to small ones: (i)  when the diameter of a channel decreases, the cell volume fraction decreases, (ii) at the same time  the value of the  optimal cell volume fraction increases.

For red blood cells, the cell flux is directly linked to the oxygen transport capacity. Interestingly, the values obtained for optimal hematocrit for vessel sizes corresponding to macrocirculation and intermediate microcirculation (arterioles) are close enough to the corresponding physiologically admitted values. Our analysis on a  network where the weight of the contribution of each vessel has been extracted from in-vivo data shows that this range of vessels also determines the RBC flow rate, indicating that the physiological values for hematocrit are close to a kind of optimum in that sense.  Our analysis also indicates the locations where active regulation processes like vasodilation or vasoconstriction are more likely to influence oxygen delivery.

Strong alterations are reported  if the reduced volume of RBCs is increased, as is known for elliptocytosis and spherocytosis diseases. Not only is the flow rate of RBCs reduced in this case compared to the flow of healthy RBCs at the same hematocrit but also the optimal hematocrit is observed to be significantly lower. The lower value of RBC flow rate within  patients suffering  these diseases implies a severe collapse of oxygen delivery, which could lead to an increased heart load in order to maintain appropriate perfusion levels. It is known that elliptocytosis and spherocytosis diseases are accompanied by a reduction of the RBC count, a consequence of the spleen filtering. Interestingly, our results show that this decrease of hematocrit probably improves oxygen delivery.

We put forward here the idea that a slightly stretchable encapsulating membrane (like polymer-based capsules)
would lead to a significant  enhancement of  oxygen transport capacity. The information generated by this study may guide the development of new soft materials, such as blood substitutes, and advance the {tuning process} and optimization of oxygen carriers. This study can also be adopted for more general questions of suspensions transport.\\

\begin{acknowledgments}
This work was partially supported by
CNES (Centre National d'Etudes Spatiales) and by the French-German university programme "Living Fluids" (grant CFDA-Q1-14). C.M. thanks CNRST (project FINCOME).
\end{acknowledgments}


%

\end{document}